# Training Gamble leads to Corporate Grumble?

David Chadwick, EuSpRIG Chair
D.R.Chadwick@gre.ac.uk

*Fifteen years of research studies have concluded unanimously that spreadsheet errors are both common and non-trivial. Now we must seek ways to reduce spreadsheet errors. Several approaches have been suggested, some of which are promising and others, while appealing because they are easy to do, are not likely to be effective. To date, only one technique, cell-by-cell code inspection, has been demonstrated to be effective. We need to conduct further research to determine the degree to which other techniques can reduce spreadsheet errors.*

As Ray Panko [1] correctly points out, the problem of spreadsheet errors has yet to be solved and represents a great, often unrecognised, risk to corporate decision making and financial integrity. Corporations are indeed gambling with the hidden risks. Abstracts of papers presented at past EuSpRIG symposia are available on the website www.eusprig.org ; a quick review of the titles suggests that the spreadsheet risk problem may be dealt with in several ways by:

- expanding awareness of the problem amongst corporations and academia Rajalingham et al [2], Ayalew et al [4], Butler R [6],Cleary P M [8]
- setting standards for end-user spreadsheet development Butler R [11] , Hawker A [3], Knight D [17]
- creating better audit approaches with accompanying software tools, Nixon D [18], Hock et al [7], Ettema H et al [16]
- Creating methodologies and software based tools for spreadsheet building Rajalingham et al [5] [14], Chadwick et al [10], O'Beirne [12], Paine [13], Raffensperger [15].

Many of the above are already being well researched and, no doubt, useful results will be obtained in due course but there is a view that the very least that can be done immediately is to concentrate effort in the first two areas mentioned. Expanding awareness and setting standards at an early enough stage in the training of would-be spreadsheet developers must surely have an effect in tackling the problem where it really begins - in human error. It is a belief held by the author that, often overlooked, is the need for better training and education of young professionals and students learning spreadsheet skills for the first time and surely destined to become the spreadsheet builders of tomorrow.

Many novice spreadsheet builders formally learn their skills by studying either for a professional examination of the auditing, IT and accountancy bodies or from courses provided at universities, colleges and private training agencies. If there is to be a true change in the quality of business spreadsheets then surely it is through such training and education providers that improvement and innovation must be channelled. The question, therefore, that needs to be addressed by EuSpRIG members is:

*'How can EuSpRIG be actively involved in the teaching and assessment of spreadsheet skills where these occur?'*

To help in answering this question it is useful to look at two areas where the influence of an interest group composed of active academics and professionals (such as EuSpRIG) may have an effect:

- the syllabi of professional examinations and coverage of spreadsheet risk,
- the curricula of higher education courses and coverage of spreadsheet risk.

## 1.0 Professional Examinations And Their Coverage Of Spreadsheet Risk

### 1.1 Professional Examinations

In the UK, there are several bodies either directly involved in the teaching and assessment of spreadsheet skills through their own professional studies or indirectly involved in the validating and influencing of syllabi of other educational providers. For instance, the ICAEW (Institute of Chartered Accountants England and Wales) is the main body influencing standards of assessment for accountants, the BCS (British Computer Society) is involved in setting its own examinations or in validating courses run at universities for IT professionals. So, too, the Institute of Internal Auditors (IIA) operates its own examinations for computer auditors. In terms of gaining a professional qualification in computer auditing and risk management there are probably two main avenues: the Certificate in Information Systems Auditing (CISA) and the Qualification in Computer Auditing (QiCA). Their approach to the development of skills differs. CISA appears suited primarily to those who have already gained some working knowledge by virtue of having done the practitioner job for some time. It results in a multiple-choice test which basically examines the applicant's in-depth knowledge of practical scenarios. It is therefore not suited to a beginner who has no or relatively limited personal experience. The QiCA, however, from the Institute of Internal Auditors, is comprised of two papers at different levels; the lower level paper is based upon learning from written texts and is more suited to the true novice, the practitioner with little experience or the student at university who both need to grasp fundamentals. The paper at the higher level deals with practical scenarios and is more appropriate for the experienced practitioner.

### 1.2 Training Curricula And Coverage Of Spreadsheet Risk

Neither the syllabus for CISA nor that for QiCA adequately cover the treatment of spreadsheet risks. A standard text used for teaching Computer auditing in the UK has one paragraph of four lines devoted to spreadsheet errors:

*'Users of spreadsheets and database packages are necessarily immune from errors. 9%ile it is true that such packages provide a safe processing environment within which it is difficult if not impossible to make undetected or obscure input and output errors, it is still possible to make errors in logic. In fact, such errors maybe more difficult to detect than they would have been if a procedural programming language had been employed... This may be the case if a very large spread sheet is generated, so that only small parts of it can be viewed on screen at any one time...* Chambers A.D and Court LM [9]

The third line admits that the problem for spreadsheets may be even worse than for normal software particularly with logical errors. Despite this, there is no literature on how to prevent errors nor on how to conduct a spreadsheet audit.

Perhaps this is where EuSpRIG has a role to play. Now established as the world's foremost action group on spreadsheet risks isn't it about time that EuSpRIG set itself up as a standards making body, a pressure group to encourage the professional bodies to incorporate spreadsheet audit in their examinations

## 2.0 Higher Education Courses And Their Coverage Of Spreadsheet Risk

## 2.1 Universities Providing Courses Recognised By Professional Bodies

The professional bodies themselves provide distance learning courses or other arrangements for study but there are several universities around the UK who offer courses preparing students either to sit for the professional audit examinations directly or to gain exemption from the same by passing other recognised courses. City Business School, Guild Hall University, University of Central England and Sheffield Hallam University are but a few of those that provide courses preparing candidates to sit the two papers of the IIA's Qualification in Computer Auditing.

Other UK universities, such as Southampton Institute and University of Greenwich, take a different approach. They have applied to the IIA for recognition of their existing undergraduate computing courses covering auditing material at the lower level of the QiCA syllabus and requesting that these courses be recognised as equivalent to the IIA's own professional exam. At the University of Greenwich the IIA has permitted students passing two particular undergraduate courses to apply for exemption from IIA studies at the lower level, an arrangement has been extremely beneficial for the teaching of computer auditing at the university.

## 2.2 Spreadsheet Teaching At Universities

There is no doubt that spreadsheet teaching at universities suffers from an image problem:

- spreadsheets are seen as trivial encompassing simple accounting models,

- the packages are easy to learn therefore the intellectual content is limited,

- students learn basic skills prior to university and have developed bad habits.

But perhaps the major problem is that of convincing academics that spreadsheet teaching requires a complete re-think involving not only changes in teaching content but also changes in teaching method perhaps accompanied by the development of a research ethos to find appropriate and innovative solutions.

## 2.3 Encouraging A Research Ethos

Possible benefits of developing a research ethos are shown below (Fig.1).

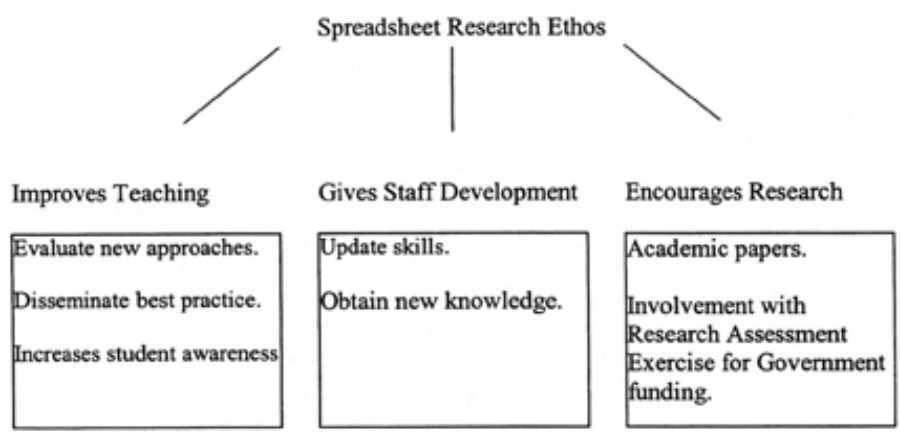

**Fig. 1 Rationale For Developing A Research Ethos**

The rationale for developing a spreadsheet research ethos is that it encourages a Research-Teaching Feedback Loop (Fig.2). Immediate improvements occur in feedback from research to classroom and feedback from classroom to research with the overall effect being an improvement in both research and teaching leading to an improvement in student learning.

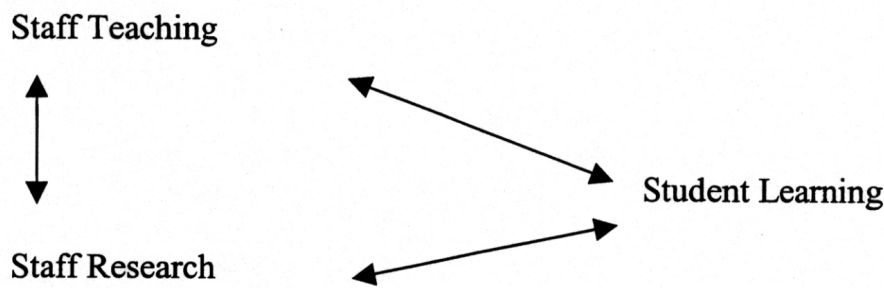

**Fig 2: Research-Teaching Feedback Loop**

The above approach necessitates a close liaison between teaching and research which in many universities tend to operate independently of each other. What is required is an approach whereby the teachers and the researchers work together to improve both the taught content (i.e. syllabus, the actual substance of imparted knowledge) and also the teaching method (i.e. types of materials, teaching style, organization of the student cohort).

**2.4 Integrating Research And Teaching**

*Recent research has highlighted the high incidence of errors in spreadsheet models used in industry. In an attempt to reduce the incidence of such errors, a teaching approach has been devised which aids students to reduce their likelihood of making common errors during development. The approach comprises of spreadsheet checking methods based on the commonly accepted educational paradigms of peer assessment and self - assessment. However, these paradigms are here based upon practical techniques commonly used by the internal audit junction such as peer audit and control and risk self - assessment. The result of this symbiosis between educational assessment and professional audit is a method that educates students in a set of structured, transferable skills for spreadsheet error checking which are useful for increasing error-awareness in the classroom and for reducing business risk in the workplace*

Chadwick D, Sue R [10]

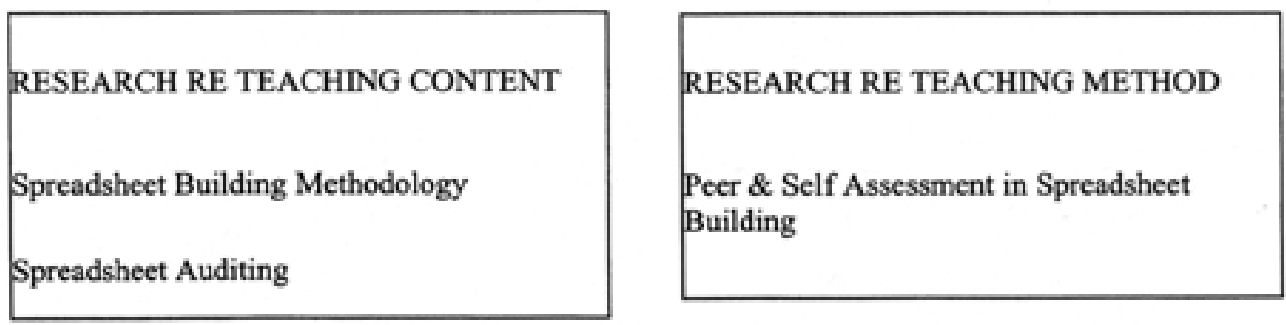

**Fig 3. Two Research Strands**

The University of Greenwich has developed its own model (Fig.3) on integrating teaching content and teaching method and also on integrating educational research with formal software engineering research with regard to spreadsheet risks (Fig. 4). In the latter there is classroom observation of students and the errors they produce whilst building spreadsheets. These findings are fed into the formal software engineering research process which is currently working on a software engineering methodology for developing spreadsheets as pieces of software. As the formal research progresses so it spins off new classroom techniques which are tried and tested and feedback given to the formal research. The classroom experience itself produces valuable findings for formal research within the arena of education.

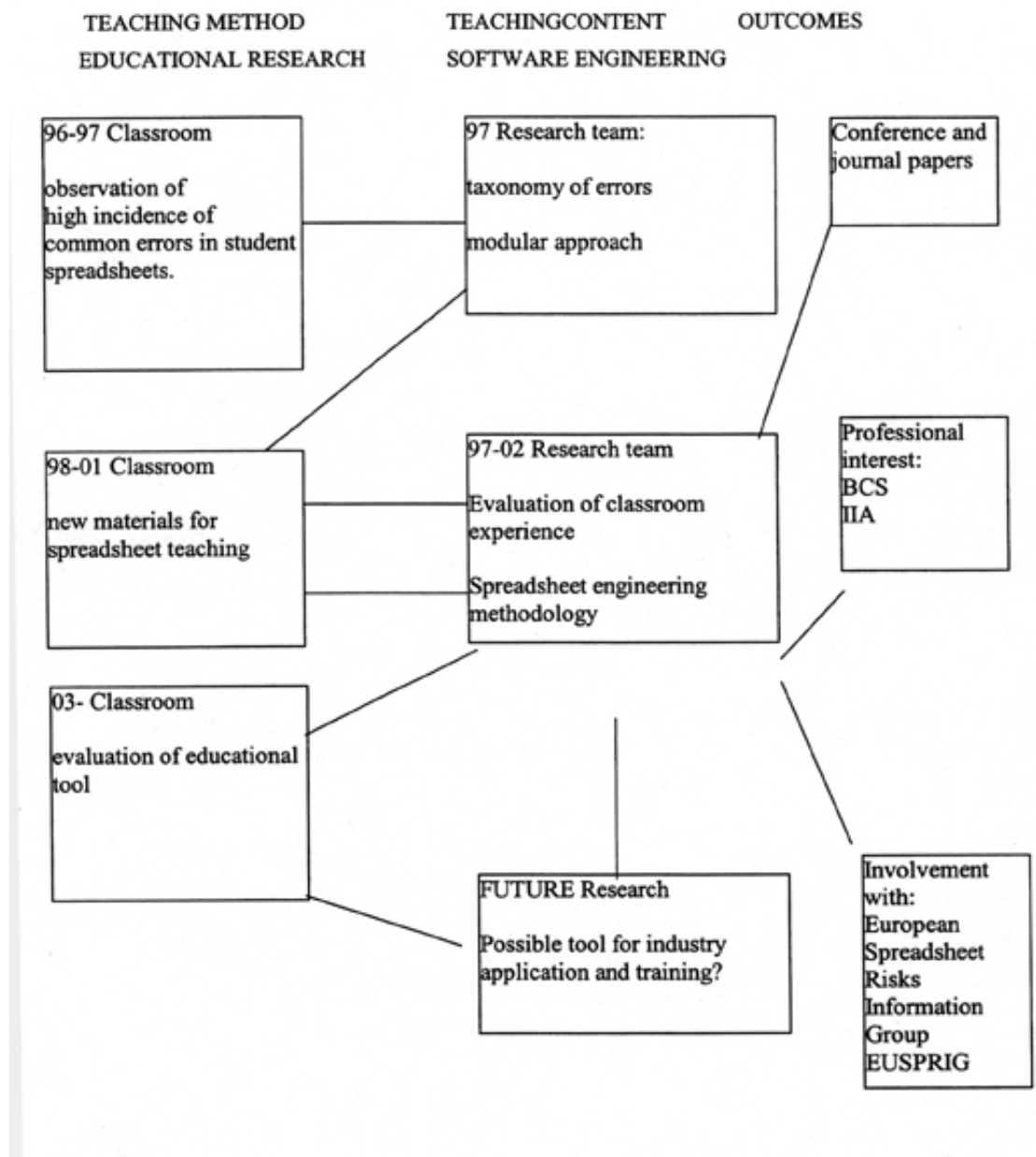

**Fig 4: The University of Greenwich model showing integration of educational and software engineering research**

## 3.0 The Importance of Standards In Training Regimes

So far we have addressed the methods of training offered by professional bodies and educational institutions. It is necessary now to look at what needs to be taught and the question of providing standards. Standards concerning spreadsheet development and audit are very important because without them, nobody would be clear just exactly what needs to be taught and how it is to be assessed. This topic generally needs more thought and debate than was possible in the preparation of this paper but suffice it to say that the matter has been brought to the attention of EuSpRIG delegates by Ray Butler at the Amsterdam symposium.

> *One of the problems reported by researchers and auditors in the field of spreadsheet risks is that of getting and keeping management's attention to the problem. Since 1996, the Information Systems Audit & Control Foundation and the IT Governance Institute have published CobiT ® which brings mainstream IT control issues into the corporate governance arena. This paper illustrates how spreadsheet risk and control issues can be mapped onto the CobiT framework and thus brought to managers' attention in a familiar format.* Butler R [11]

There is a need for the widespread promotion and eventual adoption of sound spreadsheet building and spreadsheet auditing standards. From a look at the training and teaching curricula of academia and higher education, there appears to be no attempt to present students with working standards for spreadsheet developments nor even an acknowledgement that such may be required. It seems pertinent therefore that EuSpRIG should approach these bodies and suggest that firstly, more emphasis be placed upon spreadsheet errors and that secondly, good practice techniques be promoted wherever possible. In due course perhaps EuSpRIG itself could suggest tried and tested guidelines of its own. The CobiT framework seems a useful starting point for EuSpRIG endeavours in this area.

## 4.0 How can EuSpRIG Be Involved In Education and Training?

Finally, let us consider where EuSpRIG can go from here. As an interest group composed of professional and academic contacts and with worldwide affiliations it may be able to play a decisive role in encouraging educational agencies to change their approach to spreadsheet skills training, education and assessment. At a minimum and with little effort initially EuSpRIG could be become more involved with:

- promoting a greater emphasis to be placed upon spreadsheet teaching,
- promoting research through encouragement of participation in EuSpRIG ,
- using its influence to seek funds for research-active EuSpRIG members.

## 5.0 Conclusions

The crux of this paper has been to look at perhaps the most straightforward and immediate ways in which EuSpRIG can grow and contribute directly to the problem of spreadsheet risks.

EuSpRIG has a role to play:
- in the development of syllabi in professional studies,
- in curriculum development and research in higher education,
- in the creation and promotion of spreadsheet building and auditing standards.